# Microstructure and properties of Cu-Sn-Zn-TiO$_2$ Nano-composite coatings on mild steel

Weidong Gao[1], Di Cao[1], Yunxue Jin[1], Xiaowei Zhou[1], Guang Cheng[2*], Yuxin Wang[1*]

[1] School of Materials Science and Engineering, Jiangsu University of Science and Technology, Zhenjiang, 212003, Jiangsu, China

[2] Physical and Computational Sciences Directorate, Pacific Northwest National Laboratory, P.O. Box 999, Richland, WA, 99352, USA

**Abstract:** Cu-Sn-Zn coatings have been widely used in industry for their unique properties, such as good conductivity, high corrosion resistance and excellent solderability. To further improve the mechanical performance of Cu-Sn-Zn coatings, powder-enhanced method was applied in the current study and Cu-Sn-Zn-TiO$_2$ nano-composite coatings with different TiO$_2$ concentration were fabricated. The microstructure of Cu-Sn-Zn-TiO$_2$ nano-composite coatings were investigated by X-ray diffraction (XRD) and Scanning Electron Microscopy (SEM). The mechanical properties of coatings including microhardness and wear resistance were studied. The results indicate that the incorporation of TiO$_2$ nanoparticle can significantly influence the properties of Cu-Sn-Zn coatings. The microhardness of Cu-Sn-Zn coating was increased to 383 HV from 330 HV with 1g/L TiO$_2$ addition. Also, the corrosion resistance of coating was enhanced. The effects of TiO$_2$ nanoparticle concentration on the microstructure, mechanical properties and corrosion resistance of Cu-Sn-Zn-TiO$_2$ nano-composite coatings were discussed.

---

[*] Corresponding authors.
Email address: ywan943@163.com (Yuxin Wang) septem88@hotmail.com (Guang Cheng).

# 1. Introduction

Copper-tin-zinc (Cu-Sn-Zn) ternary alloy has attracted tremendous research interests due to its unique properties, including gorgeous color, high corrosion resistance [1, 2], remarkable solder ability [3-6] and good electrical conductivity. Cu-Sn-Zn alloys have been fabricated by different methods, such as melting [7-11], powder metallurgy [12] and electrodeposition [13-16]. Among these methods, electrodeposition is an effective method to fabricate Cu-Sn-Zn coatings [17, 18] because of its simple process and high environment adaptability. In addition, the properties and color of coatings can be easily controlled by adjusting the plating parameters. The plating bath of Cu-Sn-Zn coatings mainly includes alkaline cyanide solution and acid sulphate solution. The cyanide solution is the most common bath used in the electronic applications, while the environmentally-friendly acid sulphate bath is widely adopted for decorative purpose.

Comparing with the Chromium (Cr) [19] or Nickel (Ni) [20, 21], which are the mostly used coating, Cu-Sn-Zn coating possesses a relatively soft hardness. In order to improve the mechanical properties and explore the application of Cu-Sn-Zn coatings, the preparation of Cu-Sn-Zn based nano-composite coatings, i.e., incorporating the second phase or other nanoparticles into coating matrix, could be a potential solution [20-23]. It is well known that if nano-sized particles such as oxides, carbides and polytetrafluoroethylene highly disperse into Cu-Sn-Zn coating matrix, the properties especially mechanical strength will be improved dramatically. The corresponding strengthening mechanism can be attributed to the grain refinement and dispersion strengthening.

An ideal Cu-Sn-Zn coating prepared for decoration field should have a long service life so both wear resistance and corrosion resistance should meet certain design requirements. Among various reinforcement choices, $TiO_2$ was selected due to its encouraging effect on microstructural, hardness, wear resistance and corrosion resistance of Ni-P coatings [21]. Furthermore, the preparation of $TiO_2$ added Cu-Sn-Zn coatings via environmentally-friendly acid sulphate bath have not been studied. In the present research, a systematic study of Cu-Sn-Zn-TiO$_2$ nano-composite coatings (with a TiO$_2$ concentration of 0-5 g/L) has been conducted. The influences of TiO$_2$ nanoparticles concentration on the microstructure, mechanical properties and corrosion resistance were discussed.

## 2. Experimental details

## 2.1 Sample Preparation

All the coatings were electroplated onto the hot-dipping galvanized mild steel substrates (0.2×20×30mm$^3$). The substrates were first dipped into a HCL solution (>10%) for 5 mins at room temperature to remove the covered Zn layer. The specimens were electroplated after rinse thoroughly by distilled water.

The electroplating system contains a mild steel sample as the cathode and a Cu plate as the anode. The bath composition and plating parameters are given in Table 1. The basic Cu-Sn-Zn plating bath was prepared using Sigma analytical grade reagents. The PH value of bath was adjusted to 8.6 (±0.2) by H$_2$SO$_4$ or KOH. TiO$_2$ powder with an average diameter of 10 nm was added into the electrolyte bath. The bath solution was continuously stirred at a rate of 300 rpm for 20 min by magnetic stirring in order to achieve a good suspension. Six different concentrations of TiO$_2$ powders between 0 and 5 g/L were selected in order to achieve the best properties. For easy description, Cu-Sn-Zn-x TiO$_2$ is adopted to represent the coatings containing different TiO$_2$. Pure Cu-Sn-Zn coatings were also prepared by direct deposition under the same conditions for comparison purpose.

## 2.2 Sample Characterization

The phase structure and preferred orientation of coatings were determined by X-ray diffraction (XRD) patterns using Bruker D2 Phaser diffractometer (V = 30 kV, I = 15 mA) with the Cu Kα radiation (λ = 0.15406 nm). Diffraction patterns were recorded in the 2θ range from 20º to 90º at a scanning rate of 1◦ min$^{-1}$. The morphology of as-deposited surface and composition of coatings were analyzed using a Zeiss field emission scanning electron microscope (Gemini SEM 300, Zeiss, Germany). The samples were cold mounted using Epoxy resin (ration of resin: hardener, 3:1) to obtain the cross-section images. The backscattered electrons were analyzed with a solid state detector to examine the interface between the substrate and coatings.

Vickers microhardness tests were conducted using a load of 50 g with a holding time of 15 s. The hardness values were calculated based on 10 measurements. Each indents were carried out on the surface with sufficient space between indents to eliminate the residual stress. The indent will cover a large region, i.e., over 20 $\mu m^2$ so the average properties within a large region will be obtained from microhardness test and the indentation surface roughness could be ignored because of the indentation depth [24-27]. The tribological properties of the coatings were tested using a micro-tribometer (UMT-2, CETR, USA) in air at room temperature, relative humidity of ~50% and under dry, non-lubricated conditions. The samples were tested linearly and a steel ball with a diameter of 9.58 mm was used as the counter surface. The load and total sliding distance of each wear test were set to 3N and 3m, respectively. The sliding wear test was conducted three times for each sample. The wear track images of coatings were observed by a high-resolution optical microscope.

The corrosion resistance of coatings were measured by an electrochemical workstation (CS2350, CorrTest, China). All corrosion tests were performed at room temperature (25°C) using a three-electrode system with platinum mesh as auxiliary electrode, Ag/AgCl/KCl (sat.) electrode (0.197 Vs the Standard Hydrogen Electrode-SHE) with Luggin probe as the reference electrode and coated specimen as working electrode. The exposed surface area of all samples was set to 1 $cm^2$. The potentiodynamic polarization tests were conducted at a constant scanning speed of 1 mV $s^{-1}$. The corrosion current density and corrosion potential were determined based on Tafel's extrapolation. All the corrosion tests were performed at ambient temperature in 3.5 wt.% NaCl solution.

## 3. Results and discussion

### 3.1 Phase Structure of Coatings

The X-ray diffraction patterns of Cu-Sn-Zn-x TiO$_2$ (x=0, 1, 2, 3, 4 and 5 g/L) nano-composite coatings are shown in Fig. 1. It can be observed that all the XRD patterns imply four prominent peaks which corresponding to the Cu$_{0.6}$Sn$_{0.1}$Zn$_{0.3}$ according to the phase identification. The peaks at 25.4° can be assigned to (101) plane of anatase, indicating that the TiO$_2$ nanoparticle was incorporated into the coatings. There was no TiO$_2$ peaks can be detected in Cu-Sn-Zn-1g/L TiO$_2$ and Cu-Sn-Zn-2g/L TiO$_2$ coatings due to the low quantity of TiO$_2$ particles and high intensity of other diffraction peaks.

### 3.2 Surface Morphology and Cross-Section of Coatings

Fig. 2 presents the surface morphologies of Cu-Sn-Zn-TiO$_2$ nano-composite coatings with different powder addition. The surface of pure Cu-Sn-Zn coating is relatively plain and homogeneous, as shown in Fig. 2(a). No remarkable variation can be observed on the surface of Cu-Sn-Zn-1g/L TiO$_2$ coating except some small nodules, as shown in Fig. 2(b). Further increasing TiO$_2$ concentration leads to the increase of black spots and blurry areas on the coating surface. Observation of higher magnification image confirm that the black spots depict particle agglomeration and the blurry areas represent voids in the coating. As the size of TiO$_2$ nanoparticles used for this research is very tiny (~10nm), serious TiO$_2$ particle agglomeration occurred in the Cu-Sn-Zn coating matrix, as shown in Fig. 2(b)-(f).

Fig. 3 depicts the cross-section microstructure of Cu-Sn-Zn and Cu-Sn-Zn-1g/L TiO$_2$ nano-composite coatings. A gleaming boundary between the coating and the steel substrate can be observed. No abruption or cracks exist at the interfaces of the coatings, indicating a good adhesion between the steel substrate and coating. All coatings present a similar thickness of approximately 15 μm. The deposition rate of coatings remains constant after adding TiO$_2$ nanoparticle. No obvious TiO$_2$ nanoparticles were seen in the cross-section which probably due to their small size and relatively low content. This phenomenon is consistent with the XRD analysis as shown in Fig. 1 and Fig. 2(b).

## 3.3 Microhardness and Wear Resistance of Coatings

Fig. 4 shows the correlation between the microhardness of Cu-Sn-Zn-$TiO_2$ nano-composite coatings and $TiO_2$ concentration in the bath. Initially, the microhardness of pure Cu-Sn-Zn coating was ~330 $HV_{50}$, much higher than the hardness of mild steel substrate, ~110 $HV_{50}$. Also, the small hardness deviation indicates a homogenized microstructure was prepared in the current method. With the 1g/L addition of $TiO_2$ nanoparticles to the bath, the hardness of Cu-Sn-Zn-1g/L $TiO_2$ coating reaches to the peak value, ~382 $HV_{50}$. Compared to the initial Cu-Sn-Zn coating, the hardness increased ~16%. The improvement of microhardness was attributed to the highly dispersed $TiO_2$ nanoparticles in the Cu-Sn-Zn matrix as shown in Fig. 2b and 3b. The introduced nanoparticles will increase the local dislocation density and refine the grain size. Also, the pinning effect of these nanoparticles will block the movement of the dislocations so as to raise the yield strength or the hardness of the coatings [28, 29]. However, higher $TiO_2$ concentration in the bath did not raise the hardness of the coatings. The microhardness of the coating was reduced to ~333 $HV_{50}$, ~325 $HV_{50}$, and ~320 $HV_{50}$ with 2g/L $TiO_2$, 3g/L $TiO_2$, and 4g/L $TiO_2$, respectively. The microhardness was reduced to ~300 $HV_{50}$ for Cu-Sn-Zn-5g/L $TiO_2$ coatings. Too many particles will easily lead to agglomeration and voids between particles were generated in the deposition processing as the highlighted spots shown in Fig. 2 (c), (d), (e), and (f). Those voids could significantly reduce the hardness. Hence, the competition between particles dispersion hardening and possible voids among particles determine the final hardness of the nano-composite coatings.

The wear resistance abilities of the three nano-composite coatings were tested and compared. In the dry sliding wear test, a higher hardness value brought a smaller contact area between coatings and the steel ball. So different wear tracks, 391 µm, 265 µm, and 516 µm, were observed in Cu-Sn-Zn, Cu-Sn-Zn-1g/L $TiO_2$, and Cu-Sn-Zn-5g/L $TiO_2$, respectively, as illustrated in Fig.5. The correlation between the width of wear track and hardness is quite clear. Meanwhile, different surface morphologies after wear test are observed for the three coatings because of different microstructures. Significant plowing lines and several wear debris on the worn surface were observed in Fig.5 (a). The matrix of the current Cu-Sn-Zn coating is dominant by $Cu_{0.6}Sn_{0.1}Zn_{0.3}$, which would be quite stiff. So the main wear mechanism of Cu-Sn-Zn coating is adhesive wear, which is quite common among electroplated metallic coatings [30-32]. After adding 1g/L $TiO_2$ sol into the plating solution, the contact area in the wear test reduced a lot and the plough-lines became much shallower and more uniform than the Cu-Sn-Zn coating. In addition, the wear debris was not observed in Fig.5 (b). The different surface morphologies demonstrated the reinforcement effect of nanoparticle: the embedded $TiO_2$ nanoparticles would polish the frictional surface, and homogenize the contract stress distribution between the frictional counter bodies during the wear process. With more $TiO_2$ sol into the plating solution, the voids among particle clusters raised the contact area very much. Thicker and deeper plowing lines were observed in Fig.5 (c) compared to other two samples. Also, plenty of wear debris were observed on the sliding surface.

Then, the wear volume loss of three coatings was summarized in Fig. 6. As expectation, the wear volume loss fraction for Cu-Sn-Zn-1g/L $TiO_2$ is ~0.7 ×$10^{-13}$ $m^3$, which is the smallest among tested samples. The volume loss of Cu-Sn-Zn and Cu-Sn-Zn-5g/L $TiO_2$ are ~ 2.4 ×$10^{-13}$ $m^3$ and ~ 7.2 ×$10^{-13}$ $m^3$, respectively. According to the Archard's law, the volume loss during sliding wear is inversely proportional to the material hardness and proportional to the friction coefficient [33-36]. Because of the similar surface roughness in the current study, the friction coefficient of different coatings are quite similar. Thus, the volume loss among different coatings would directly reflect the differences between hardness. It should be noted that the Vickers hardness test would penetrate the coating-substrate system for over 5 μm under 50 g load which has already exceeded the 10% thickness of the coating [23, 37, 38]. On this condition, the soft mild steel substrate has affected the hardness values. The actual hardness difference between Cu-Sn-Zn-1g/L TiO$_2$ and Cu-Sn-Zn should be much more than 16% if the substrate effect was removed. Hence, indentation study with penetration depth less than 1 μm will be conducted in the future.

## 3.4 Corrosion Resistance of Coatings

Fig. 7 presents the polarization curves of Cu-Sn-Zn, Cu-Sn-Zn-1g/L TiO$_2$ and Cu-Sn-Zn-5g/L TiO$_2$ nano-composite coatings. The corresponding electrochemical properties, including corrosion potentials, corrosion current densities and corrosion rates of coatings were calculated by fitting potentiodynamic curves as listed in Table 2.

The corrosion potentials ($E_{corr}$) of three coatings were approximately the same, around -0.256 V. The corrosion current densities ($I_{corr}$) and corrosion rates of coatings varied with the concentration of TiO$_2$ addition. The $I_{corr}$ and corrosion rate of Cu-Sn-Zn coating were ~ 8.353 μA/cm$^2$ and 0.098 mm/a, while the $I_{corr}$ and corrosion rate value obtained for Cu-Sn-Zn-1g/L TiO$_2$ coating were ~ 6.284 μA/cm$^2$ and 0.074 mm/a, respectively, indicating an improved corrosion resistance. The improvement of corrosion resistance could be attributed to the dense structure of coating and the uniform distribution of TiO$_2$ nanoparticles in coating matrix. However, more TiO$_2$ nanopowder addition will cause a deterioration of corrosion resistance for the coatings due to the serious particle agglomeration and porous microstructure. Cu-Sn-Zn-5g/L TiO$_2$ coating exhibits highest $I_{corr}$ and the fastest corrosion rate, ~ 9.191 μA/cm$^2$ and 0.108 mm/a, respectively. The results of Tafel plots show a same tendency with the mechanical properties of coatings.

## 4. Conclusions

Cu-Sn-Zn-TiO$_2$ nano-composite coatings were produced by electrodeposition method using acid sulfate bath. Their microstructure, mechanical properties and corrosion resistance were systematically investigated. Proper inclusion of TiO$_2$ nanoparticles into Cu-Sn-Zn matrix provides a decent strengthening effect and significantly improves the properties of coatings. Compared to the pure Cu-Sn-Zn coating, the microhardness of Cu-Sn-Zn-1g/L TiO$_2$ coating was increased by ~16%. Meanwhile, the corresponding wear resistance and corrosion resistance have also been greatly enhanced. Future study will focus on developing new Cu-Sn-Zn based nano-composite coatings by reducing the agglomeration of particles in the deposition process to prepare densified coatings. Meanwhile, the composition optimization and more characterizations methods, i.e., high-resolution transmission electron microscopy (HRTEM) [22, 23], resonance ultrasound spectroscopy [39-41], etc., will be attempted to better understand its microstructure.

## Acknowledgments

This research was supported by National Natural Science Foundation of China (51601073), Jiangsu Distinguished Professor Project (1064901601) and Start research fund of Jiangsu University of Science and Technology (1624821607-5). Pacific Northwest National Laboratory (PNNL) is operated by Battelle Memorial Institute for the US Department of Energy (DOE) under Contract No. DE-AC06–76RL01830. The authors would like to express our gratitude to Mr. Haixin Zhu, Mr. Lin Zhang and technical staff in Jinshi Electroplating Ltd.

# Tables

Table 1 Bath composition and processing parameters of electrodeposited Cu-Sn-Zn-TiO$_2$ and Cu-Sn-Zn coatings.

| Bath composition & plating parameters | Quantity |
|---|---|
| CuSO$_4$ | 35 g/L |
| SnCl$_2$ | 3 g/L |
| ZnSO$_4$ | 10 g/L |
| K$_4$P$_2$O$_7$ | 260 g/L |
| NaH$_2$PO$_4$·2H$_2$O | 10 g/L |
| C$_6$H$_5$K$_3$O$_7$ | 25 g/L |
| KNaC$_4$H$_4$O$_6$·4H$_2$O | 25 g/L |
| Saccharin | 0.04 g/L |
| Temperature | 32 °C |
| Current density | 20 mA/cm$^2$ |
| Agitation speed | 300 rpm |
| Plating time | 60 min |
| TiO$_2$ nanoparticles | 0, 1g/L, 2g/L, 3 g/L, 4 g/L, 5 g/L |

Table 2 Corrosion parameters derived from analysis of potentiodynamic linear polarization curves

| Sample | Ecorr (V) | Icorr ($\mu Acm^{-2}$) | Corrosion Rate (mm/a) |
|---|---|---|---|
| Cu-Sn-Zn | -0.253 | 8.353 | 0.098 |
| Cu-Sn-Zn-1g/L TiO$_2$ | -0.256 | 6.284 | 0.074 |
| Cu-Sn-Zn-5g/L TiO$_2$ | -0.259 | 9.191 | 0.108 |

# Figures

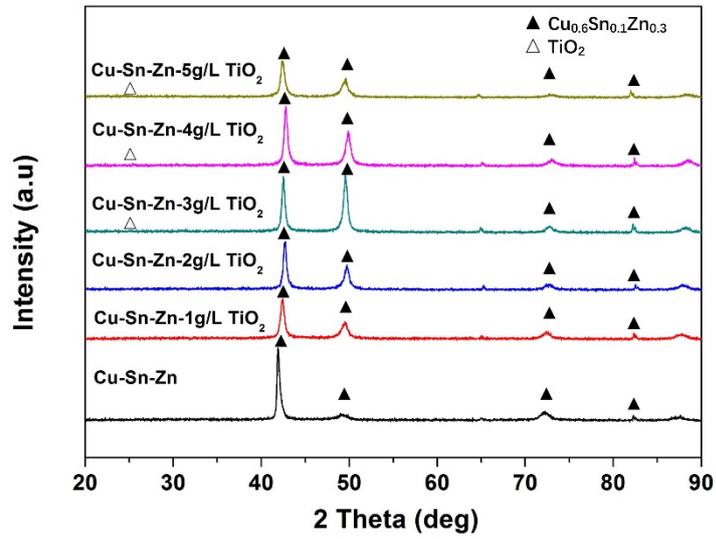

Fig. 1 XRD patterns of Cu-Sn-Zn and Cu-Sn-Zn-TiO$_2$ nano-composite coatings

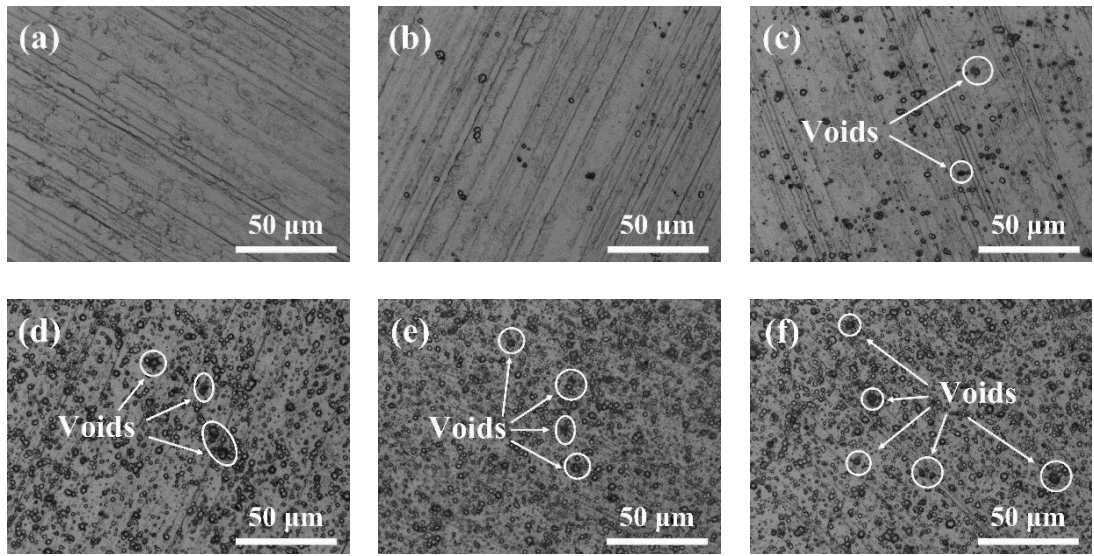

Fig. 2 Surface morphologies of six coating investigated in the current study: (a) Cu-Sn-Zn, (b) Cu-Sn-Zn-1g/L TiO$_2$, (c) Cu-Sn-Zn-2g/L TiO$_2$, (d) Cu-Sn-Zn-3g/L TiO$_2$, (e) Cu-Sn-Zn-4g/L TiO$_2$, and (f) Cu-Sn-Zn-5g/L TiO$_2$.

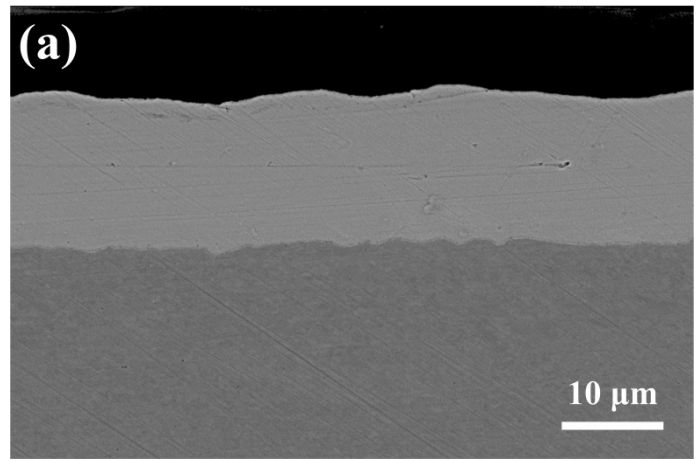
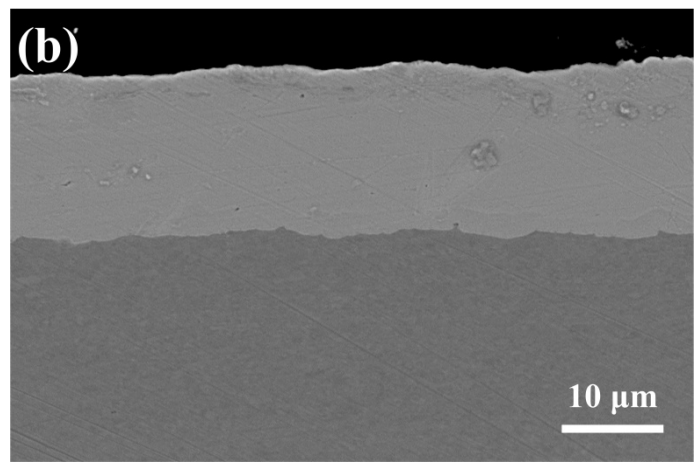

Fig. 3 Cross-section images of (a) Cu-Sn-Zn and (b) Cu-Sn-Zn-1g/L TiO$_2$ coatings

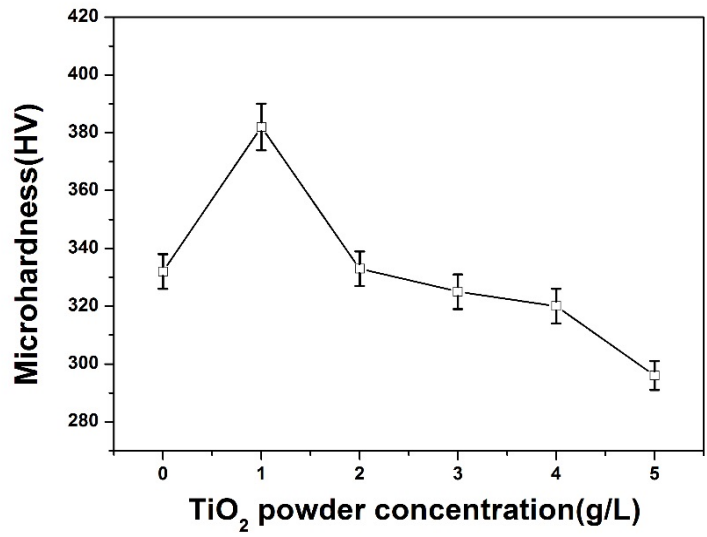

Fig. 4 Effect of TiO$_2$ concentration on microhardness of Cu-Sn-Zn-TiO$_2$ nano-composite coating.

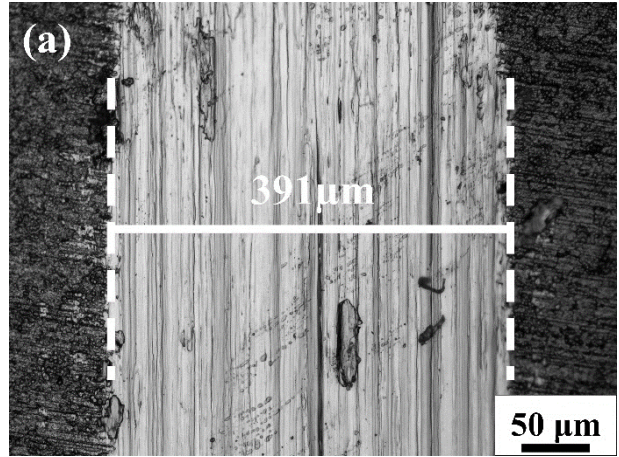

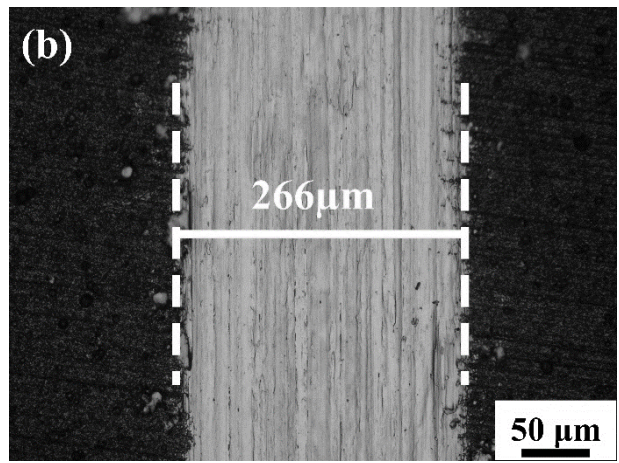

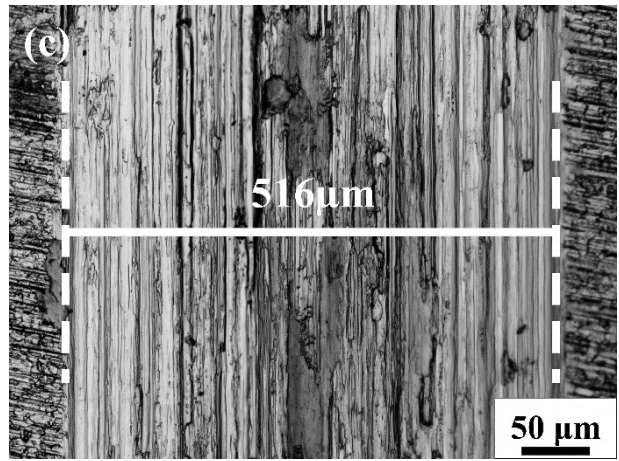

Fig. 5 Wear track images of (a) Cu-Sn-Zn, (b) Cu-Sn-Zn-1 g/L $TiO_2$ and (c) Cu-Sn-Zn-5 g/L $TiO_2$ coatings.

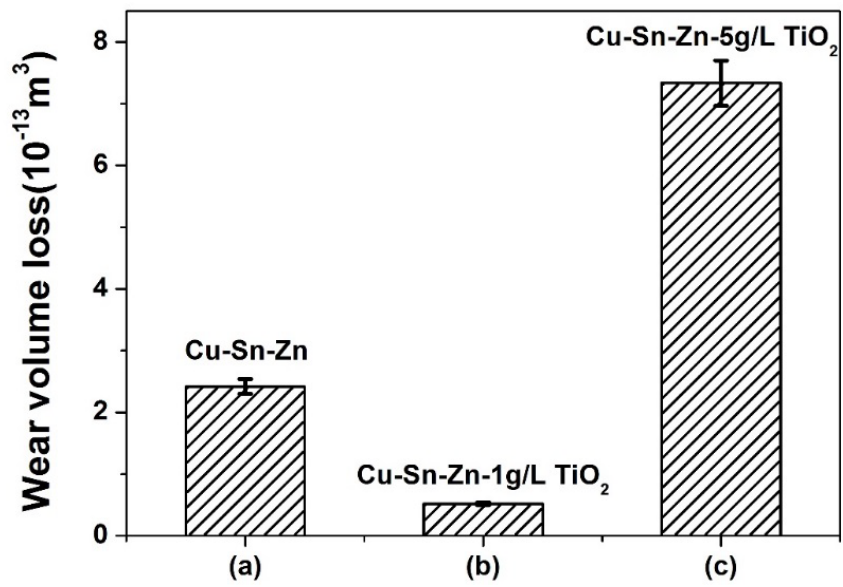

Fig. 6 Wear volume loss of (a) Cu-Sn-Zn, (b) Cu-Sn-Zn-1 g/L $TiO_2$ and (c) Cu-Sn-Zn -5 g/L $TiO_2$ coatings.

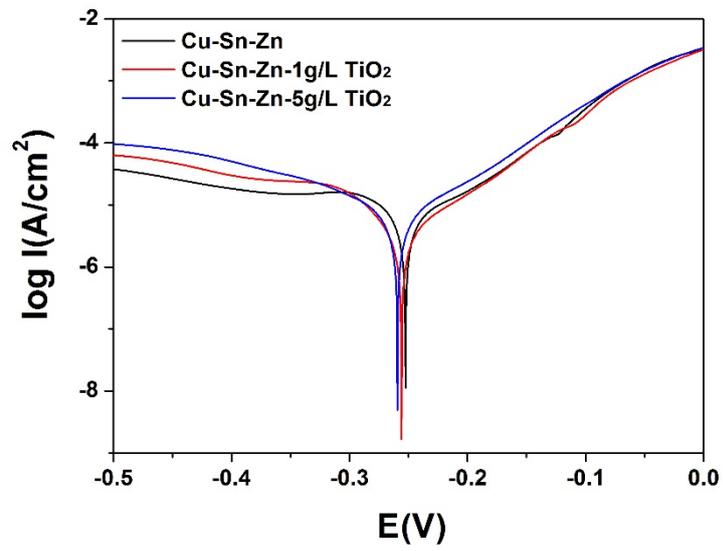

Fig. 7 Potentiodynamic polarization curves of (a) Cu-Sn-Zn, (b) Cu-Sn-Zn-1 g/L TiO$_2$ and (c) Cu-Sn-Zn -5 g/L TiO$_2$ coatings.